\documentclass[preprint]{aastex}

\newcommand{\kms}{\mbox{km s$^{-1}$}}

\newcommand{\Msun}{\mbox{$M_{\sun}$}}
\newcommand{\Lsun}{\mbox{$L_{\sun}$}}
\newcommand{\gcoldcore}{G305.136$+$0.068}
\newcommand{\cotd}{\mbox{CO$(3\rightarrow2)$}}
\newcommand{\tcotd}{\mbox{$^{13}$CO$(3\rightarrow2)$}}
\newcommand{\csdu}{\mbox{CS$(2\rightarrow1$)}}
\newcommand{\cstd}{\mbox{CS$(3\rightarrow2$)}}
\newcommand{\cscc}{\mbox{CS$(5\rightarrow4$)}}
\newcommand{\csss}{\mbox{CS$(7\rightarrow6$)}}

\newcommand{\ccgtcc}{\mbox{G305.136+0.068}}
\newcommand{\skipthis}[1]{}
\newcommand{\hii}{\mbox{\ion{H}{2}}}

\shortauthors{Garay et al.}
\shorttitle{Young massive and dense core}
\usepackage{amssymb,amsmath,natbib,graphicx}

\begin{document}



\title{G305.136+0.068: A massive and dense cold core in an
early stage of evolution}

\author{Guido Garay\altaffilmark{1}, Diego Mardones\altaffilmark{1}, 
Yanett Contreras\altaffilmark{1,2}, Jaime E. Pineda\altaffilmark{3}, 
Elise Servajean\altaffilmark{1}, and Andr\'es E. Guzm{\'{a}}n\altaffilmark{1,4}}

\altaffiltext{1}{Departamento de Astronom\'{\i}a, Universidad de Chile,
Casilla 36-D, Santiago, Chile}
\altaffiltext{2}{CSIRO Astronomy and Space Science, P.O. Box 76, Epping 
NSW 1710, Australia}
\altaffiltext{3}{Institute for Astronomy, ETH Zurich,
Wolfgang-Pauli-Strasse 27, 8093 Zurich, Switzerland}
\altaffiltext{4}{Harvard-Smithsonian Center for Astrophysiscs, 60 Garden Street,
 Cambridge, Massachusetts, USA}

\begin{abstract}

We report molecular line observations, made with ASTE and SEST, and dust continuum 
observations at 0.87 mm, made with APEX, towards the cold dust core G305.136+0.068. 
The molecular observations show that the core is isolated and roughly circularly 
symmetric and imply that it has a mass of $1.1\times10^3$ \Msun. A simultaneous 
model fitting of the spectra observed in four transitions of CS, using a 
non-LTE radiative transfer code, indicates that the core is centrally condensed, 
with the density decreasing with radius as $r^{-1.8}$, and that the turbulent 
velocity increases towards the center. The dust observations also indicate that 
the core is highly centrally condensed and that the average column density is 
1.1 g cm$^{-2}$, value slightly above the theoretical threshold required for 
the formation of high mass stars. A fit to the spectral energy distribution 
of the emission from the core indicates a dust temperature of $17\pm2$ K, 
confirming that the core is cold. {\it Spitzer} images show that the 
core is seen in silhouette from 3.6 to 24.0 $\mu$m and that is surrounded by an 
envelope of emission, presumably tracing an externally excited photo-dissociated 
region. We found two embedded sources within a region of 20\arcsec\ centered at 
the peak of the core, one of which is young, has a luminosity of 66 \Lsun\ and 
is accreting mass with a high accretion rate, of $\sim1\times10^{-4}$ \Msun\ 
yr$^{-1}$. We suggest that this object corresponds to the seed of a high mass 
protostar still in the process of formation. The present observations support 
the hypothesis that \ccgtcc\ is a massive and dense cold core in an early stage 
of evolution, in which the formation of a high mass star has just started.

\end{abstract}

\keywords{ISM: clouds --- ISM: dust --- stars: formation --- stars: massive}

\vfill\eject

\section{Introduction}

Several single dish surveys of molecular line emission in high density tracers 
\citep{Plume1992,Plume1997,Juvela1996} and of dust continuum emission 
\citep{Beuther2002,Mueller2002,Faundez2004,Williams2004,Garay2007} carried out toward luminous sources, either ultra 
compact (UC) \hii\ regions and/or luminous IRAS sources, have revealed that 
young high mass stars are usually found in massive (M$ \sim10^3$ \Msun) and 
dense ($5\times10^5$ cm$^{-3}$) cores with sizes of typically $\sim$0.4 pc and 
temperatures of typically $\sim$ 32 K. Since high mass stars have already been 
formed in these cores, which may have appreciably affected their natal 
environment through stellar winds and radiation, it is not clear whether or 
not the properties of these cores are representative of the initial conditions 
for the formation of high mass stars at the scale of parsecs.

The observational search for massive dense cold cores, capable of forming
high mass stars but in a stage before star formation begins,
started only recently, with the advent of telescopes that allow to identify 
them \citep{Garay2004, Hill2005, Sridharan2005, Schuller2009, Contreras2013}. 
The bulk of their luminosity is expected to be emitted at mm and
sub-mm wavelengths. Since high mass stars are formed in clusters, the 
determination of the initial conditions of massive star formation is  
key to understand the formation process that produces a cluster of stars.

In this paper we report observations of molecular line emission and of  
continuum emission at 0.87 mm toward G305.136+0.068, a massive dust core 
identified by \citet{Garay2004} from mm observations. The absence of 
emission in any of the MSX and IRAS bands from this core suggested it is cold, 
and therefore a bonafide candidate for being a core in an early stage of 
evolution.  Archival data from the {\it Spitzer Space Telescope} GLIMPSE survey
and from the {\it Herschel Space Observatory} Hi-GAL project were analyzed to 
search for embedded sources and to determine the dust temperature of the core, 
respectively.  

\section{Observations}

\subsection{Molecular lines}

The molecular line observations were made using the 10-m Atacama Submillimeter 
Telescope Experiment (ASTE) located in Pampa La 
Bola, Chile, and the 15-m Swedish--ESO Submillimetre Telescope (SEST) located 
on La Silla, Chile. The observed transitions and basic observational parameters 
are summarized in Table~\ref{tbl-obs}.
\begin{deluxetable}{ccccccccc}
  \tablewidth{0pt}
\tablecolumns{9}
  \tablecaption{Observational Parameters \label{tbl-obs}}
  \tablehead{
    \colhead{Line} & \colhead{Frequency} & \colhead{Tel.} & \colhead{Beam} 
    & \colhead{$\eta_{mb}$} & \colhead{Map} & \colhead{Spacing} &
     \colhead{Channel width} & \colhead{Noise} \\
   \colhead{} & \colhead{(MHz)} & \colhead{} &\colhead{(\arcsec)} 
&\colhead{} &\colhead{($\arcmin\times\arcmin$)}& \colhead{(\arcsec)} 
&\colhead{(\kms)} &\colhead{(K)}}
\startdata
\csdu &  97980.968 & SEST & 52 & 0.73  &$4\times4$ & 30 & 0.103 & 0.040 \\
\cstd & 149969.049 & SEST & 34 & 0.66  &$4\times4$ & 30 & 0.087 & 0.076 \\ 
\cscc & 244935.606 & SEST & 22 & 0.40  &$3\times3$ & 30 & 0.156 & 0.123 \\
\csss & 342882.950 & ASTE & 22 & 0.65  &$3\times3$ & 20 & 0.109 & 0.094 \\
\cotd & 345795.990 & ASTE & 22 & 0.65  &$5\times5$ & 30 & 0.108 & 0.090 \\
\tcotd & 330587.960& ASTE & 23 & 0.65  &$5\times5$ & 30 & 0.113 & 0.086 \\
\enddata
\end{deluxetable}

The ASTE observations were carried out during December 2004 and July 2005. 
A detailed description of the characteristics of ASTE is given by
\citet{Ezawa2004}.  The frontend consisted of a single pixel heterodyne
SiS receiver operating in the 345 GHz band. The backend consisted of four
digital autocorrelation spectrometers, each with 1024 spectral channels.
The half-power beam width of the telescope at 345 GHz is
$\sim22\arcsec$. The main beam efficiency, during night observations, was 0.65.
We mapped the CO(3$\rightarrow$2) and $^{13}$CO(3$\rightarrow$2)
emission within a region of $150\arcsec\times150$\arcsec\ in size, with angular
spacings of 30\arcsec, centered at the peak of the dust core. On-source
integration times per map position were 1 and 3 minutes
for CO(3$\rightarrow$2) and $^{13}$CO(3$\rightarrow$2), respectively.
System temperatures were typically $\sim 210$~K and $\sim 175$~K at 
the frequencies of the CO(3$\rightarrow$2) and $^{13}$CO(3$\rightarrow$2) 
lines, respectively. In addition, we mapped the \csss\ emission 
within a region of $60\arcsec\times 60\arcsec$ in size, with 20\arcsec\ angular 
spacing. The system temperature was typically $\sim 370$~K and 
the on-source integration time per map position was 4 minutes. 
The data were reduced with the program NEWSTAR and then converted to
fits format for further processing with CLASS. 

The SEST observations were carried out during March 2003. We
used the high resolution acousto-optical spectrometers which provided a channel
separation of 43 kHz and a total bandwidth of 43 MHz. We mapped the
CS(2$\rightarrow$1) and CS(3$\rightarrow$2) emission across a region
of 2\arcmin\ in diameter, with angular spacings of 30\arcsec, centered at the
peak of the dust core. On-source integration times per map position were 
6 and 3 minutes for the CS(2$\rightarrow$1) and CS(3$\rightarrow$2) lines,
respectively.  The CS(5$\rightarrow$4) emission was mapped across a region of
1.5\arcmin\ in diameter, with angular spacing of 30\arcsec.
On-source integration time per map position was 6 minutes.
System temperatures were typically $\sim200$~K at 3 mm, $\sim290$~K at 2 mm
and $\sim430$~K at 1 mm.

\subsection{Dust continuum}

The 0.87 mm continuum observations were made using the 12-m Atacama Pathfinder
Experiment Telescope (APEX) located at Llano de Chajnantor, Chile. As receiver 
we used the Large Array Bolometer Camera (LABOCA), which is an array made 
of 295 bolometers arranged in a hexagonal pattern, with two-beam spacing between 
bolometers 
\citep{Siringo2009}. The passband of the bolometers has an equivalent width
of $60$ GHz and is centered at $345$ GHz. The HPBW of a single element is
19.2\arcsec.  The observations were carried out during May of 2008, covering 
$-55$ to $-60$ deg in longitude, and from $-1$ to $+1$ in latitude and are 
part of the 
Apex Telescope Large Area Survey of the Galactic Plane (ATLASGAL) 
survey at 850 $\mu$m \citep{Schuller2009}. 
Off-line data reduction was performed using the software
package BoA following the steps described in \citet{Siringo2009}.  The
flux calibration was derived from maps of Uranus. We estimate that the
uncertainty in the absolute calibration and pointing accuracy are
$20\%$ and 3\arcsec, respectively. The noise level achieved in the
image is 59 mJy/beam.

\subsection{Ancillary public data}

In addition to the above observations, we made use of publicly available 
mid-infrared images taken with the {\it Spitzer Space Telescope}
and far-infrared images taken with the {\it Herschel Space Observatory}. 
Mid-infrared images at 3.6, 4.6, 5.8 and 8.0 $\mu$m were obtained with the 
Infrared Array Camera \citep[IRAC,][]{Fazio2004} as part 
of the GLIMPSE Legacy Program. Data at 24 $\mu$m were obtained 
with the Multi-band Imaging Photometer as part of the MIPSGAL 
survey \citep{Carey2006}.
The far-infrared images were taken as part of the Herschel Hi-GAL key
project \citep{Molinari2010} using the parallel fast-scanning mode and
obtained through the Herschel Science Archive. Images at 70 and 160
$\mu$m were observed using the PACS bolometer \citep{Poglitsch2010}
and images at 250, 350 and 500 $\mu$m were obtained
using the SPIRE bolometer \citep{Griffin2010}. The two orthogonal scan
directions were combined into a single map using the Herschel Interactive
Processing Environment (HIPE, v9.2). Cross-scan combination and de-striping
was performed using the standard HIPE tools over the Hi-GAL maps.


\section{Results}

\subsection{Molecular lines}

\begin{figure}
\epsscale{1.0}
\plotone{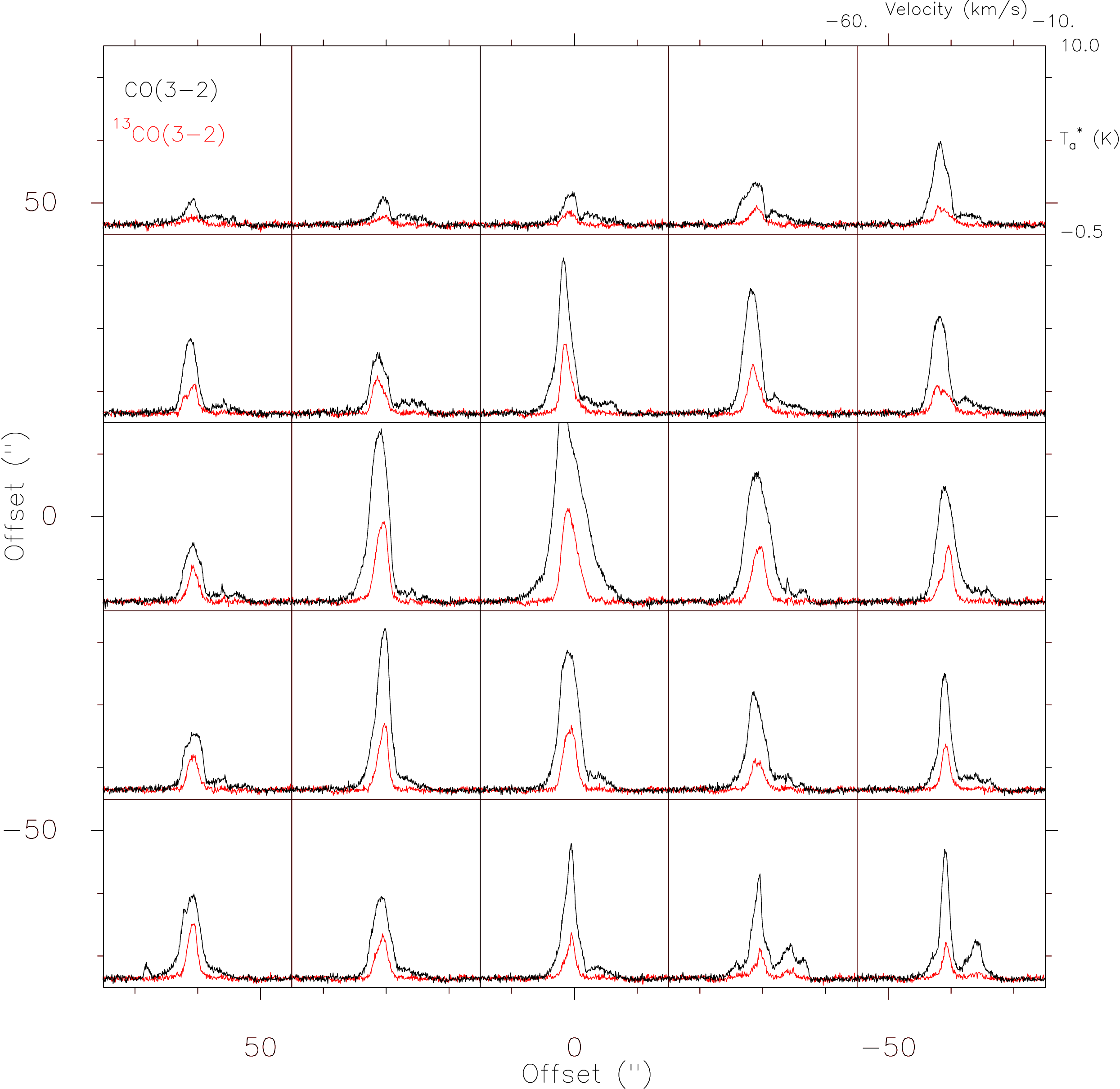}
\vspace{-5.0mm}
\caption
{Spectra of the \cotd\ and \tcotd\ line emission observed with ASTE toward 
the \ccgtcc\ core. 
The angular separation between panels is 30\arcsec.  The velocity scale
in each panel runs from $-60$ to $-10$ \kms, and the temperature
scale from $-0.5$ to 10.0 K.
\label{fig-co_spemap}}
\end{figure}

Figure~\ref{fig-co_spemap} shows the \cotd\ and \tcotd\ spectra observed with
ASTE within a 150\arcsec$\times$150\arcsec\ region, centered at the peak 
position of the dust core. In both transitions the profiles of the 
emission from the core ambient gas, centered at $-36.5$ \kms, are nearly Gaussian, 
except in the \cotd\ line at the central position.
Toward the whole observed region, the \cotd\ spectra show weak emission at 
smaller velocities than that of the ambient cloud, which most likely arises 
from foreground or background molecular clouds.

Table~\ref{tbl-linepar} gives the observed parameters of the spatially 
averaged emission in all the observed transitions. Columns 2 -- 5 give, 
respectively, the peak antenna temperature, line center velocity, line width, 
and velocity integrated antenna temperature, determined from a Gaussian fit to 
the source averaged spectra. The observed line widths are broad, typically 
$\sim 4$ \kms, much larger than the thermal width expected for a cloud with 
T$_K\sim17$ K, indicating that the \ccgtcc\ core is highly turbulent. 
At the peak position of the core the \cotd\ and \csdu\ spectra show an excess 
wing emission which may indicate the presence of outflowing gas.
\begin{deluxetable}{lccccc}
\tablewidth{0pt}
\tablecolumns{6}
\tablecaption{MOLECULAR LINES: OBSERVED PARAMETERS$^{\rm a}$ \label{tbl-linepar}}
\tablehead{
\colhead{Line}   & \colhead{T$_A^*$} & \colhead{V$_{\rm LSR}$} & 
\colhead{$\Delta$v (FWHM)} &
  \colhead{$\int{T_A^*dv}$} & \colhead{$\theta_d$}  \\ 
\colhead{} & \colhead{(K)} &  \colhead{(\kms)} &  \colhead{(\kms)} &   
  \colhead{(K~\kms)} & \colhead{(\arcsec)} 
}
\startdata
\csdu  & 0.763 & $-36.41\pm0.02$ & $4.15\pm0.04$ & $3.37\pm0.03$ & 48 \\   
\cstd  & 0.378 & $-36.52\pm0.02$ & $3.54\pm0.05$ & $1.42\pm0.02$ & 38 \\    
\cscc  & 0.156 & $-36.35\pm0.11$ & $2.59\pm0.26$ & $0.431\pm0.04$ & 31 \\ 
\csss  & 0.108 & $-36.48\pm0.20$ & $5.40\pm0.64$ & $0.618\pm0.05$ & 18 \\ 
\cotd  & 4.79  & $-36.73\pm0.01$ & $5.75\pm0.01$ & $29.3\pm0.05$ & 64 \\ 
\tcotd & 2.06  & $-36.30\pm0.01$ & $3.95\pm0.02$ & $8.66\pm0.05$ & 59 \\ 
\enddata
\tablenotetext{a}{From Gaussian fits to the spatially averaged emission.}
\end{deluxetable}

\begin{figure}
\epsscale{1.0}
\plotone{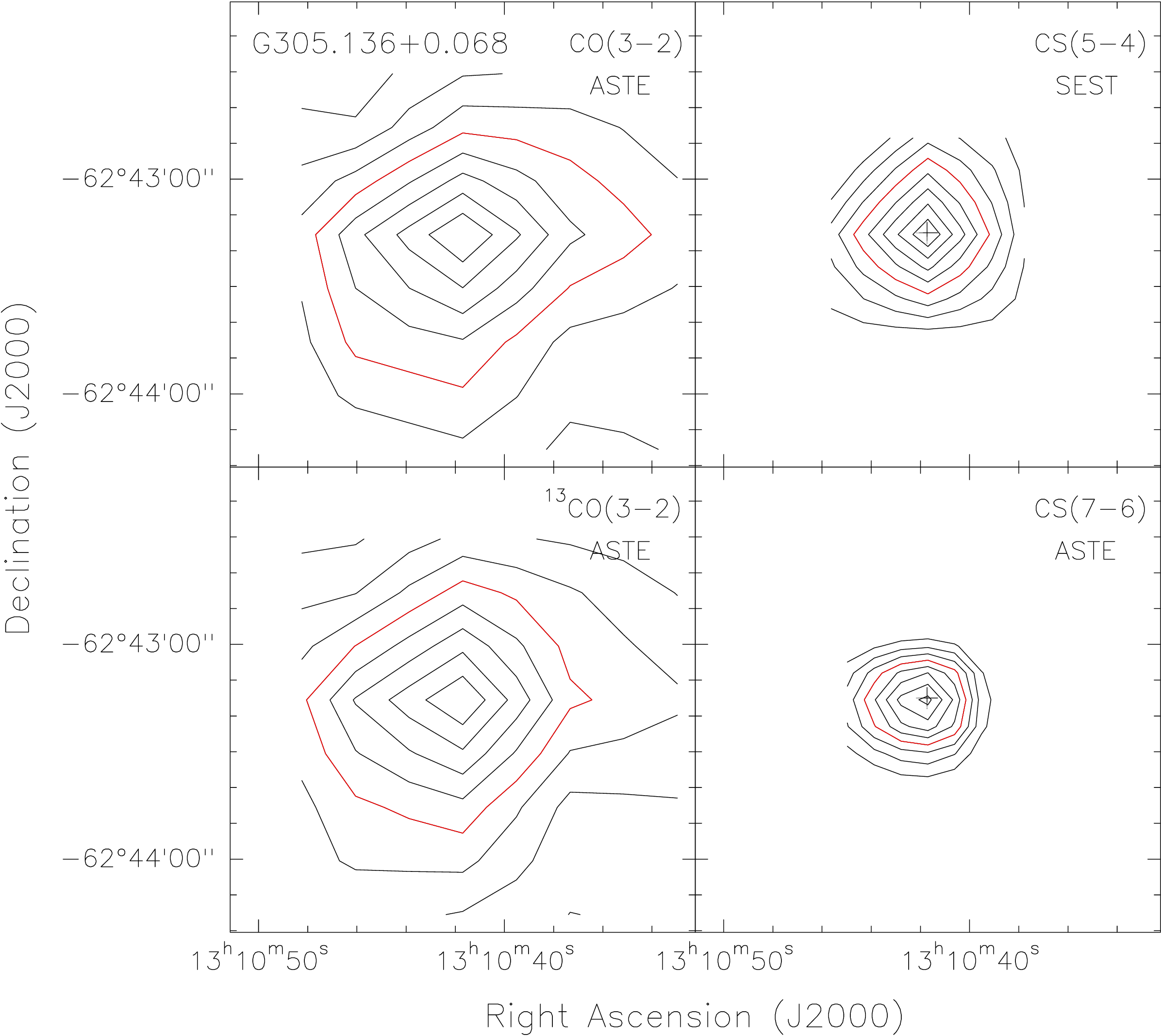}
\vspace{2.0mm}
\caption
{Contour maps of the velocity integrated ambient line emission from the 
\ccgtcc\ core. Contour levels are drawn at 20, 30, 40, 50, 60, 70, 80, and
90\% of the peak emission. Upper left: \cotd. Peak emission: 78.5 K \kms.  
Lower left: \tcotd. Peak emission: 25.0 K \kms. Upper right:
\cscc. Peak emission: 1.95 K \kms. Lower right: \csss. Peak 
emission: 2.2 K \kms. The cross marks the peak position of the dust 
continuum emission at 1.3 mm \citep{Garay2004}.  
\label{fig-maps}}
\end{figure}

Figure~\ref{fig-maps} presents contour maps of the velocity integrated 
ambient gas emission in the \cotd, \tcotd, \cscc\ and \csss\ 
lines.  The range of velocity integration is from $-40.9$ to $-31.1$ \kms\
for the CO lines and from $-40.0$ to $-33.3$ \kms\ for the CS transitions.
G305.136+0.068 appears as a distinct isolated molecular core.
Column 6 of Table~\ref{tbl-linepar} gives the FWHM angular size of the core in 
each observed transition. Assuming that the core is at a distance of 3.4 kpc 
\citep{Garay2004}, the radius in the different transitions range from 0.15 
to 0.5 pc.

\subsection{Dust emission}

\begin{figure}
\epsscale{1.0}
\plotone{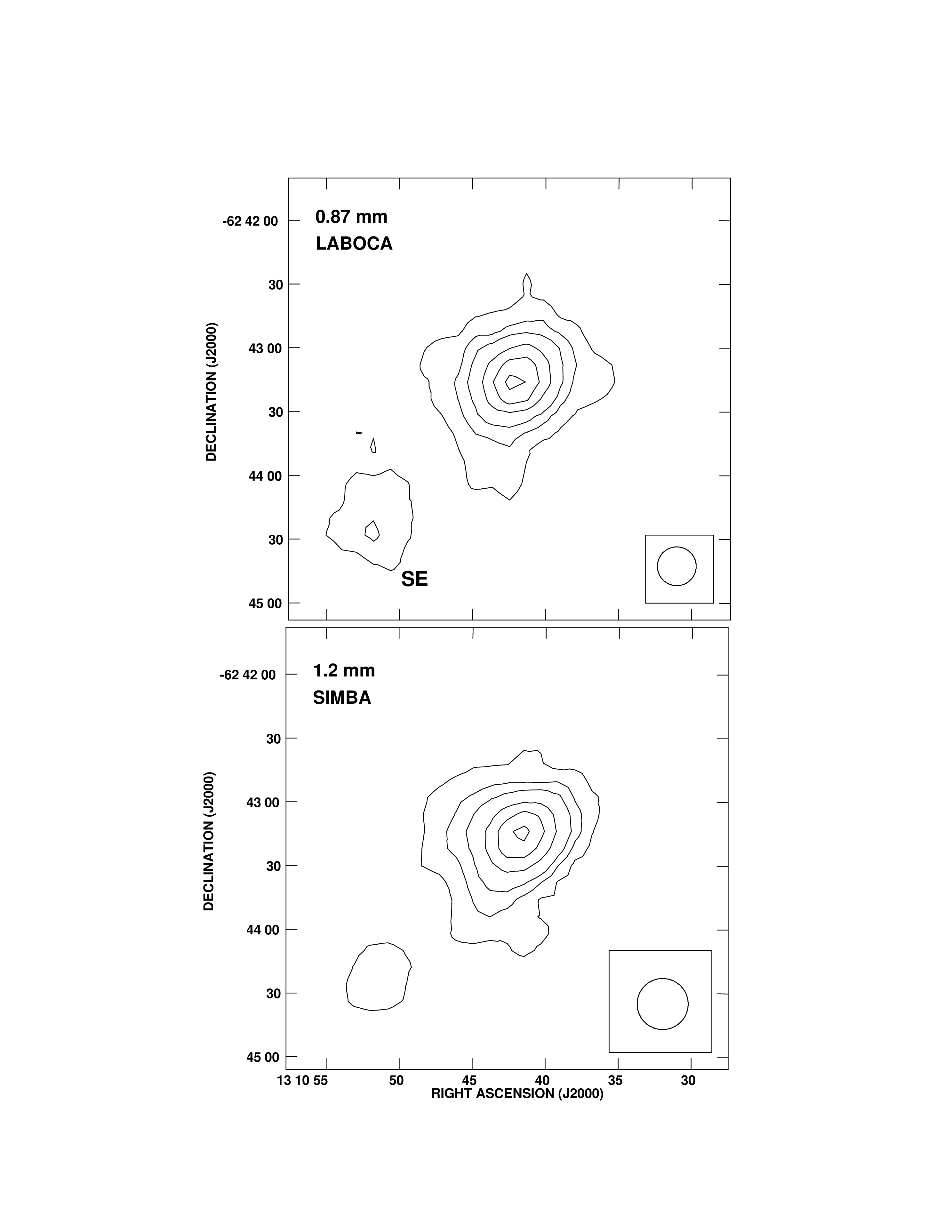}
\vspace{-20.0mm}
\caption
{\baselineskip3.0pt
Maps of dust continuum emission observed towards \gcoldcore. 
The FWHM beam is shown at the bottom right.
Top: 0.87-mm emission. Contour levels are drawn at 1 2 3 5 7 and 9 
$\times$ 0.35 Jy/beam. The 1$\sigma$ rms noise is 59 mJy/beam.
Bottom: 1.2-mm emission. Contour levels are drawn at 1 2 3 5 7 and 9 $\times$
0.15 Jy/beam. The 1$\sigma$ rms noise is 36 mJy/beam.
\label{fig-dustcont}}
\end{figure}

Figure~\ref{fig-dustcont} presents a contour map of the 0.87 mm continuum emission
observed towards the G305.136+0.068 core (upper panel) and, for comparison, a 
contour map of the 1.2 mm continuum emission (lower panel; \citet{Garay2004}). 
The extent and morphology of the dust continuum emission is similar at both 
wavelengths. The observed parameters of the cold core at 0.87 mm are given in
Table~\ref{tbl-obspardust}. Also given in this Table are the parameters of the 
weak 0.87-mm source located $\sim1.6$\arcmin\ SE of G305.136+0.06. Columns 2 and 3 
give the peak position. Columns 4 and 5 give, respectively, the peak flux 
density and the total flux density, and column 6 gives the deconvolved major and 
minor FWHM angular sizes determined by fitting a single Gaussian profile to the 
spatial distribution.  We note, however, that Gaussian fits to the spatial 
distribution of the dust continuum emission do not give good fits. 
\begin{deluxetable}{lcccccc}
\tablewidth{0pt}
\tablecolumns{7}
\tablecaption{0.87-MM CONTINUUM EMISSION: OBSERVED PARAMETERS
  \label{tbl-obspardust}}
\tablehead{
\colhead{Source}  & \multicolumn{2}{c}{Peak position} & \colhead{} &
 \multicolumn{2}{c}{Flux density\tablenotemark{a}} & \colhead{Angular size
\tablenotemark{b}} \\
\cline{2-3} \cline{5-6}
\colhead{}                  & \colhead{$\alpha$(2000)}      &
\colhead{$\delta$(2000)}  &  & \colhead{Peak}  & \colhead{Total} &
\colhead{(\arcsec)}  \\
\colhead{}                  & \colhead{h~~m~~s~~}      &
\colhead{$^{\circ}~~\arcmin~~\arcsec$} &   & \colhead{(Jy/beam)}  & \colhead{(Jy)} &
\colhead{} 
}
\startdata
G305.136+0.068   & 13 10 42.06 & -62 43 15.6  & & 3.37 & 16.1 &  $39\times31$ \\
G305.136+0.068SE & 13 10 51.97 & -62 44 24.2  & & 0.76 &  3.2 &  $39\times33$ \\
\enddata
\tablenotetext{a}{Errors in the flux density are dominated by the
uncertainties in the flux calibration, of $\sim$15\%.}
\tablenotetext{b}{Errors in the angular sizes are typically 10\%.}
\end{deluxetable}

\begin{figure}
\epsscale{1.0}
\plotone{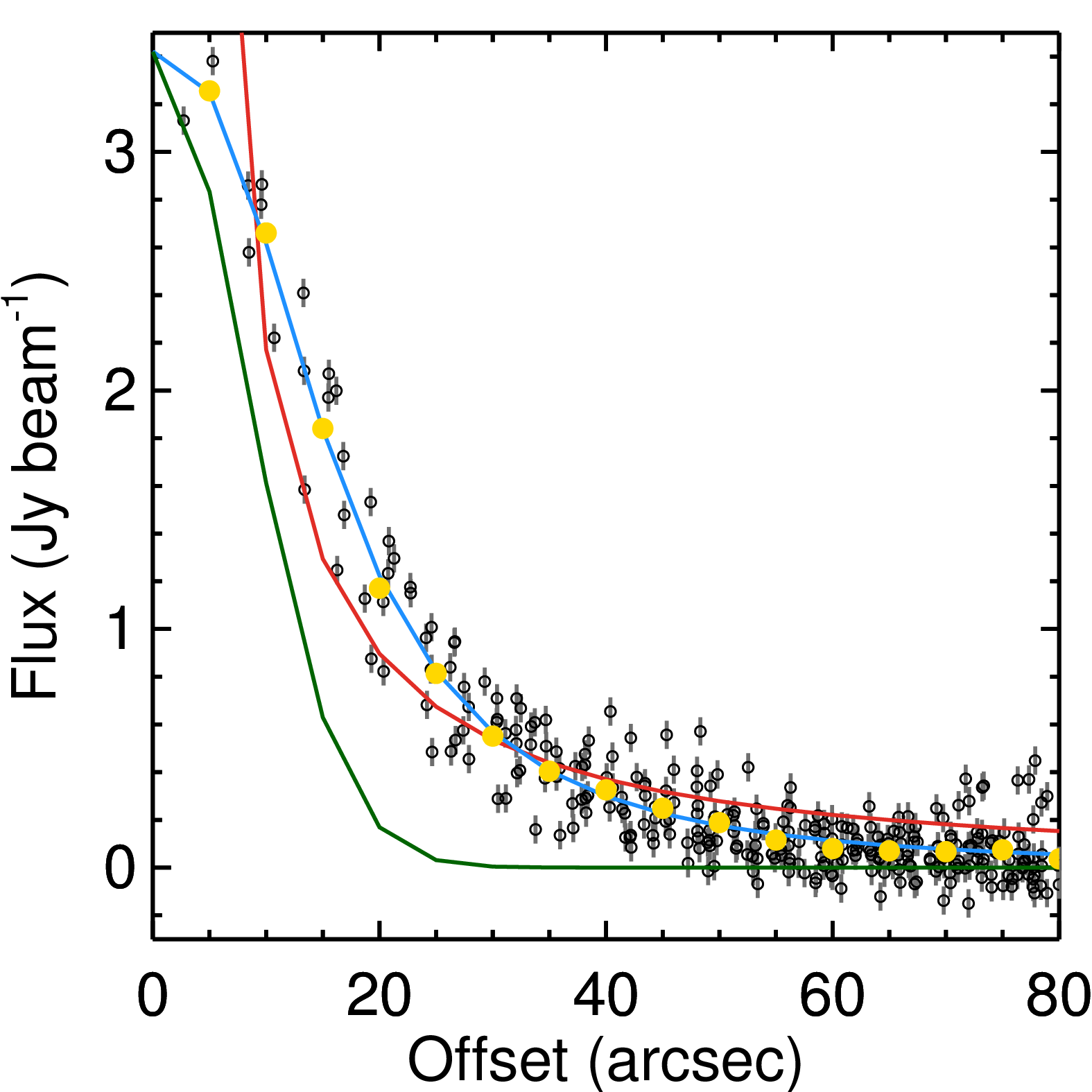}
\vspace{-10.0mm}
\caption
{\baselineskip3.0pt
Radial intensity profile of the 0.87-mm continuum emission towards the 
\ccgtcc\ core. The blue and red line corresponds, respectively, to a fit of 
the observed radial intensity with the function given in eqn.(1) and a 
power law fit. The yellow circles show the average flux at different annuli.
The green line corresponds to the beam response of the telescope at 0.87-mm
 \citep{Siringo2009}.
\label{fig-dustprofile}}
\end{figure}

\begin{figure}
\epsscale{1.0}
\plotone{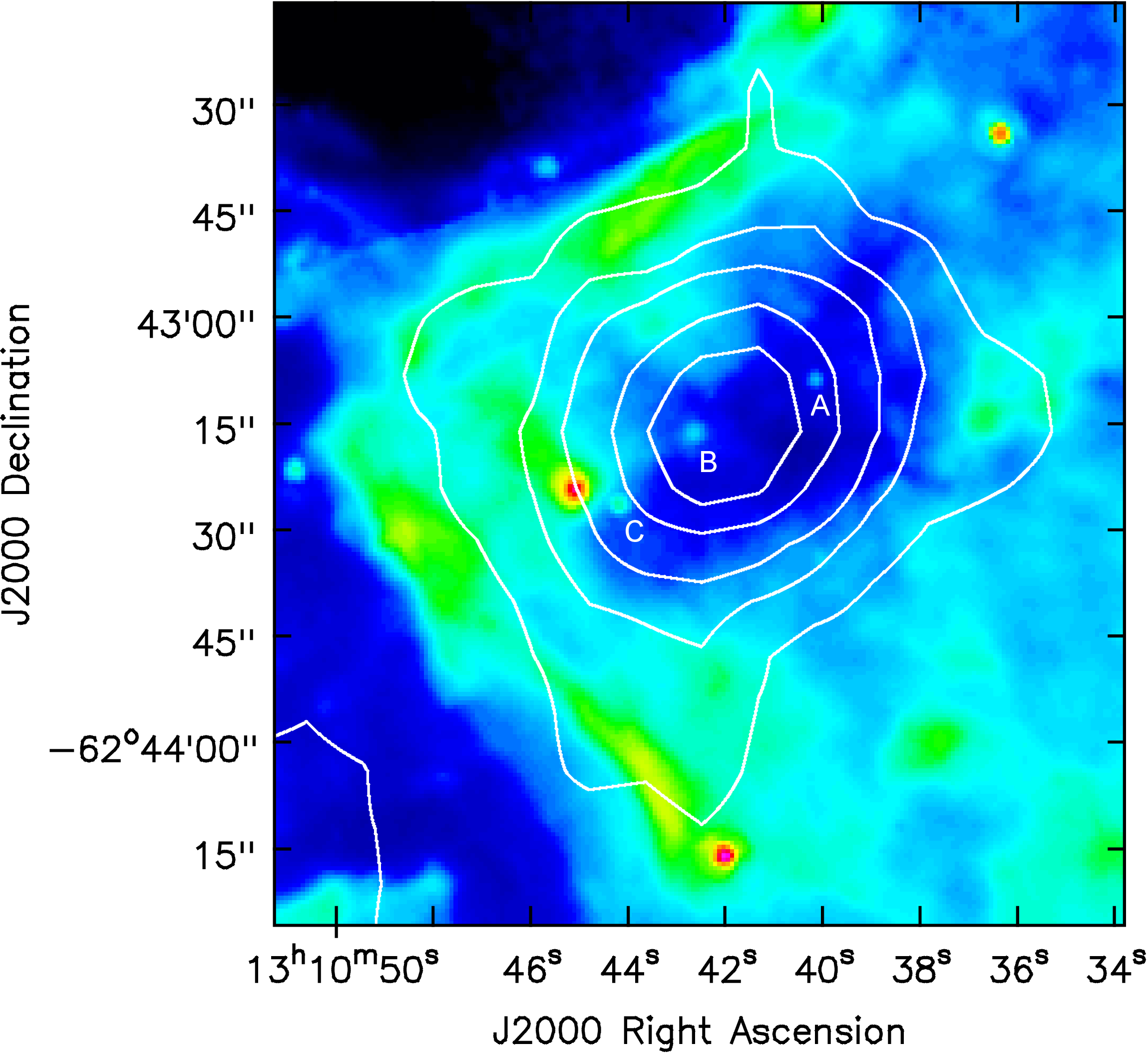}
\vspace{5.0mm}
\caption
{\baselineskip2.0pt
8.0 $\mu$m GLIMPSE image of the \ccgtcc\ core
overlaid with contours of the 0.87 mm emission. Labeled are 
compact sources projected toward the dust core. 
\label{fig-irac4}}
\end{figure}

Figure~\ref{fig-dustprofile} shows the observed flux density per beam at 0.87 mm
as a function of radial distance from the peak position of the core, within a 
region of 
$80\arcsec$ radius. The yellow circles show the average flux at uniformly
spaced annuli of 5\arcsec\ width. A power-law fit to the radial intensity profile, 
shown by the red curve,
does not provide a good fit to the central and outer regions. We find that the 
observed radial intensity profile, $I(r)$, is best reproduced with the 
function,
\begin{equation}
I(r) = \frac{I_0}{1+(r/r_{0})^p}~~,
\label{eqn-iprofile}
\end{equation}
where $r$ is the distance from the center, $p$ is the asymptotic power-law exponent,
$I_0$ is the central intensity, and $r_0$ is the radius of the central ``flat'' 
region. The best fit, shown by the blue curve in Figure~\ref{fig-dustprofile},
gives the following parameters $r_0=15.9\pm0.2$ \arcsec,
$I_0=3.42\pm0.05$\,Jy\,beam$^{-1}$, and $p=2.54\pm0.03$.
If we assume that the temperature and opacity does not change 
much as a function of radius then $I(r) \propto N(r)$, where $N(r)$
is the column density, suggesting that the core has a steep 
density profile.  \citet{Tafalla2004}, using the same parametrization, 
found similar density profiles for starless low-mass dense cores.
The mid-infrared GLIMPSE and MIPS images show that the molecular core 
is seen in silhoutte against the Galactic background emission, indicating 
that is cold and that has opacities greater than 1 at all wavelengths 
shorter than $24~\mu$m.  
Furthermore, in all bands the core appears to be surrounded by an 
envelope of emission, with an angular radius of $\sim$45". This is 
illustrated in Figure~\ref{fig-irac4} which presents the 8.0 $\mu$m GLIMPSE 
image image towards the \ccgtcc\ core.
The envelope is most prominent in the 5.8 and 8.0 $\mu$m images 
and is weak in the 4.5 $\mu$m image. The 5.8 and 8.0 $\mu$m bands are sensitive to 
emission from polycyclic aromatic hydrocarbon (PAH) molecules, whereas the 
4.5 $\mu$m band does not contain any PAH emission feature.
Thus, the envelope emission is most likely to arise from a photo-dissociated 
region externally heated by the Galactic UV radiation field.

\section{Discussion}

\subsection{Parameters derived from molecular line observations}

Optical depths, column densities and total mass of the core were 
derived from the observations of the CO and $^{13}$CO lines as follows. 
From the ratio of the observed brightness temperatures in the CO and 
$^{13}$CO lines, we derived the opacity in the $^{13}$CO line as a 
function of the velocity in each of the observed positions of the core
(see expressions [A2] and [A3] of \citet{Bourke1997}. We assumed an 
excitation temperature of 18 K and a [CO/$^{13}$CO] abundance ratio of 50.

The total column density of the $^{13}$CO molecule, in each of the observed 
positions, was then computed from the opacity in the \tcotd\ line, $\tau_{13}$,  
and excitation temperature, $T_{ex}$, using the expression 
\citep[e.g.,][]{Garden1991} 
\begin{equation}
N(^{13}{\rm CO}) =  0.81\times10^{14}
\frac{(T_{ex}+0.88)\exp(15.87/T_{ex})}{1-\exp(-15.87/T_{ex})} \int \tau_{13} dv ~~
{\rm cm}^{-2}.
\label{eqn-n13co}
\end{equation}
where $v$ is measured in \kms. This expression assumes that all energy 
levels are populated according to local thermodynamic equilibrium at 
the temperature $T_{ex}$.
We find that the $^{13}$CO column density at the peak position of the 
core is $2.0\times10^{16}$ cm$^{-2}$. Integrating the column densities 
over the solid angle subtended by the core and assuming a 
[H$_2$/$^{13}$CO] abundance ratio of $5\times10^5$ \citep{Pineda2008},
we derive that the core has 
a total mass of $1.1\times 10^3 M_\odot$. The molecular density derived from 
this mass and the observed radius in the $^{13}{\rm CO}$ line, of 0.49 pc, 
assuming that the core has uniform density, is $\sim 4\times10^4$ cm$^{-3}$.

Another estimate of the total mass using the molecular line observations can be 
made assuming that the core is in virial equilibrium \citep{MacLaren1988}. From 
the observed size and average line width 
in the \tcotd\ and \csdu\ transitions, we derive virial masses of $1.4\times10^{3}$ 
and $1.2\times10^{3}$ \Msun, respectively.

\subsection{Parameters derived from dust observations}

\begin{figure}
\epsscale{1.0}
\plotone{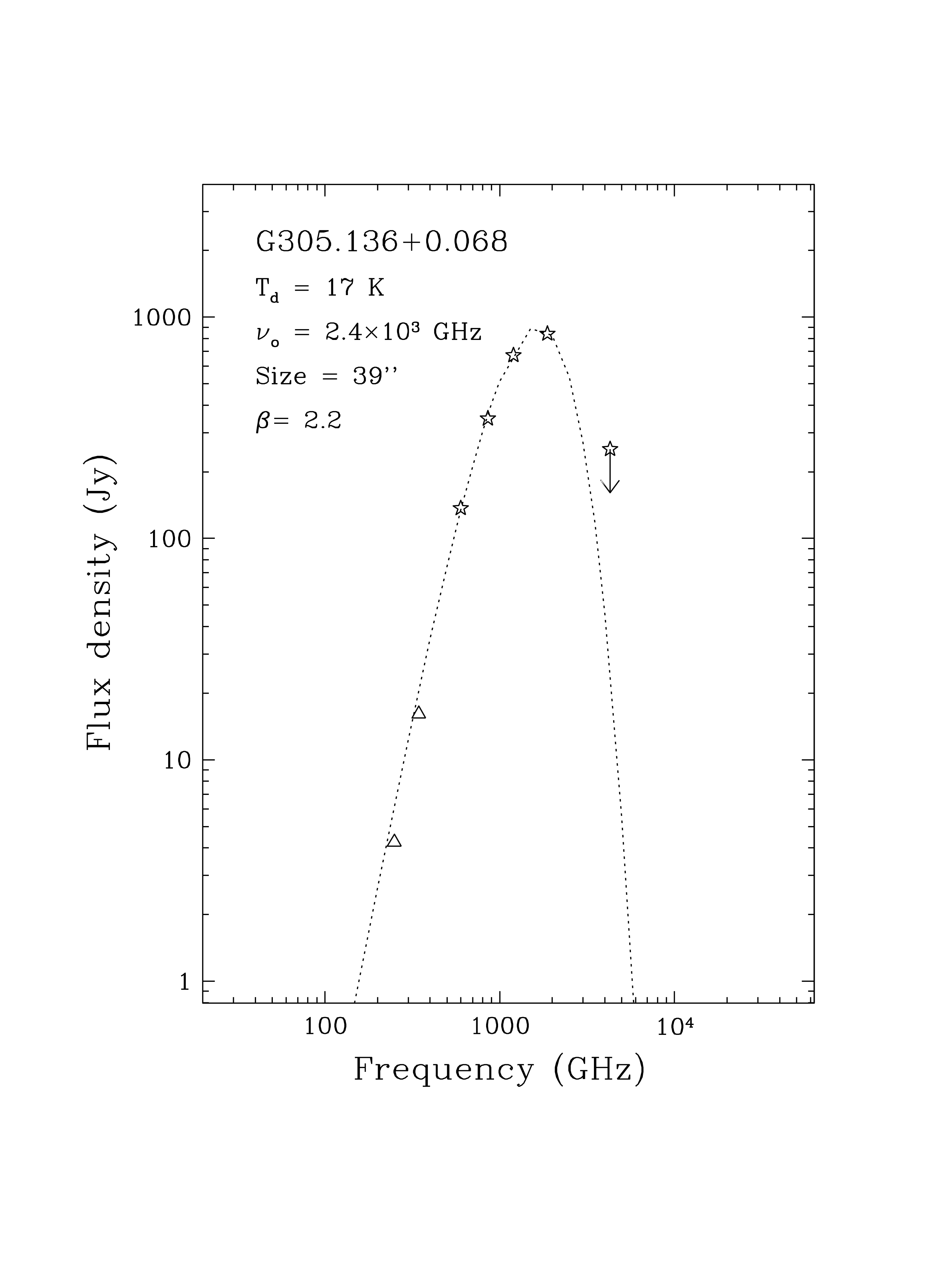}
\vspace{-35.0mm}
\caption
{\baselineskip3.0pt
Spectral energy distribution of the \gcoldcore\ core. The dotted line 
indicates the best fit with a single temperature modified
black body model. Listed are the fitted parameters. 
\label{fig-sed}}
\end{figure}

The parameters of the core derived from the 0.87 mm observations are 
summarized in Table~\ref{tbl-deriveddust}. To determine the 
core dust temperature we made use of the publicly available Herschel 
images in the wavelength range from 70 to 500 $\mu$m. Figure~\ref{fig-sed} 
shows the spectral energy distribution (SED) from 70 $\mu$m to 1.2 mm. In 
this wavelength range the emission is mainly due to thermal dust emission.
The flux density at 70~\micron\ correspond to an upper limit.
We fitted the SED using a modified blackbody function of the form
$B_{\nu}(T_d)\left[1-\exp(-\tau_{\nu})\right]\Omega_s~, $
where $\tau_{\nu}$ is the dust optical depth, $B_{\nu}(T_d)$ is the Planck
function at the dust temperature $T_d$, and $\Omega_s$ is the solid angle 
subtended by the dust emitting region. The opacity was assumed to vary with 
frequency as $\nu^{\beta}$, i.e. $\tau_{\nu}= \left(\nu/\nu_o\right)^{\beta}$, 
where $\nu_o$ is the frequency at which the optical depth is unity, 
refered as the turnover frequency. 
The best fit to the SED, shown as dotted line in Figure~\ref{fig-sed}, indicates 
a dust temperature of 17 K, a turnover frequency of $2.4\times10^3$ GHz, an 
angular size of 39$\arcsec$ and a dust opacity power law index of 2.2.
The column density derived from the turnover frequency is 
$2.6\times10^{23}$ cm$^{-2}$. This value, equivalent to 1.1 g cm$^{-2}$, is 
slightly above the theoretical threshold required for the formation
of high mass stars \citep{Krumholz2008}, suggesting that the \ccgtcc\ 
core will form high mass stars. 
\begin{deluxetable}{lcccccc}
\tablewidth{0pt}
\tablecolumns{7}
\tablecaption{DUST EMISSION: DERIVED PARAMETERS \label{tbl-deriveddust}}
\tablehead{
\colhead{Source} & \colhead{T$_d$} & \colhead{Radius} &
 \colhead{Mass} &  \colhead{n(H$_2$)}  & \colhead{N(H$_2$)} &
 \colhead{$\tau_{0.87mm}$} \\
 \colhead{} & \colhead{(K)} & \colhead{(pc)} & \colhead{(M$_{\odot}$)} &
 \colhead{(cm$^{-3})$} & \colhead{(cm$^{-2}$)}  & \colhead{}
}
\startdata
G305.136+0.068 &  17. & 0.29 & $1.2\times10^3$ & $2.0\times10^5$
& $2.4\times10^{23}$ & 0.02 \\
\enddata
\end{deluxetable}

Assuming that the observed flux density at 0.87 mm, $S_{0.87mm}$,
corresponds to optically thin, isothermal dust emission, then
the mass of the core, given in Column 4 of Table~\ref{tbl-deriveddust}, was 
determined using the expression \citep[e.g.~][]{Chini1987},
\begin{equation}
M_{g} = \frac{S_{0.87mm} D^2}{R_{dg} \kappa_{0.87mm} B_{0.87mm}(T_d)} ~~,
\label{eqn-mdust}
\end{equation}
where $R_{dg}$ is the dust-to-gas mass ratio (assuming 10\% He) with a 
value of $0.01$, $\kappa_{0.87mm}$ is the mass absorption coefficient of 
dust with a value of 2 cm$^2$ g$^{-1}$ \citep{Ossenkopf1994}, and $B(T_d)$ 
is the Planck function at the dust temperature $T_d$.  The mass derived
is $1.2\times10^3$ \Msun, in good agreement with the LTE mass and
the virial masses.  The average molecular density and average
column density, derived from the mass and
radius assuming that the core has uniform density, are given in columns 
5 and 6 of Table~\ref{tbl-deriveddust}.  We stress that these values should
be taken with caution, since as discussed in \S 3.2, the
\gcoldcore\ cold core exhibits a steep density gradient.
Finally, column 7 of Table~\ref{tbl-deriveddust} gives the
average continuum optical depth at 0.87 mm. It takes a value of
0.02, consistent with the optically thin assumption.

The similar values of the molecular gas mass
(derived from both the dust continuum emission and LTE analysis)
and the virial mass (which measures the gas plus stellar mass), suggest that 
a significant fraction of the total mass of this massive dense cold core is 
in the form of molecular gas, and hence that the gas dominates the gravitational 
potential.

\subsection{Parameters derived from line profile modelling}

As previously discussed, the physical conditions in the core are unlikely to 
be uniform, but depend with the distance from the central region.
To determine the physical structure of the cloud, model profiles in the
CS~J=2$\rightarrow$1, 3$\rightarrow$2, 5$\rightarrow$4, and 7$\rightarrow$6 
lines were computed and compared with the observed profiles.
The model profiles were obtained using a Monte Carlo radiative transfer code 
\citep{Mardones1998} based on \citet{Bernes1979}.
To obtain a reliable CS~(7$\rightarrow$6) profile we used 14 levels and 13
transitions in the Monte Carlo simulation. The robustness of the Monte Carlo method
is widely discussed by van \citet{vanZadelhoff2002}.
\begin{figure}
\epsscale{1.0}
\plotone{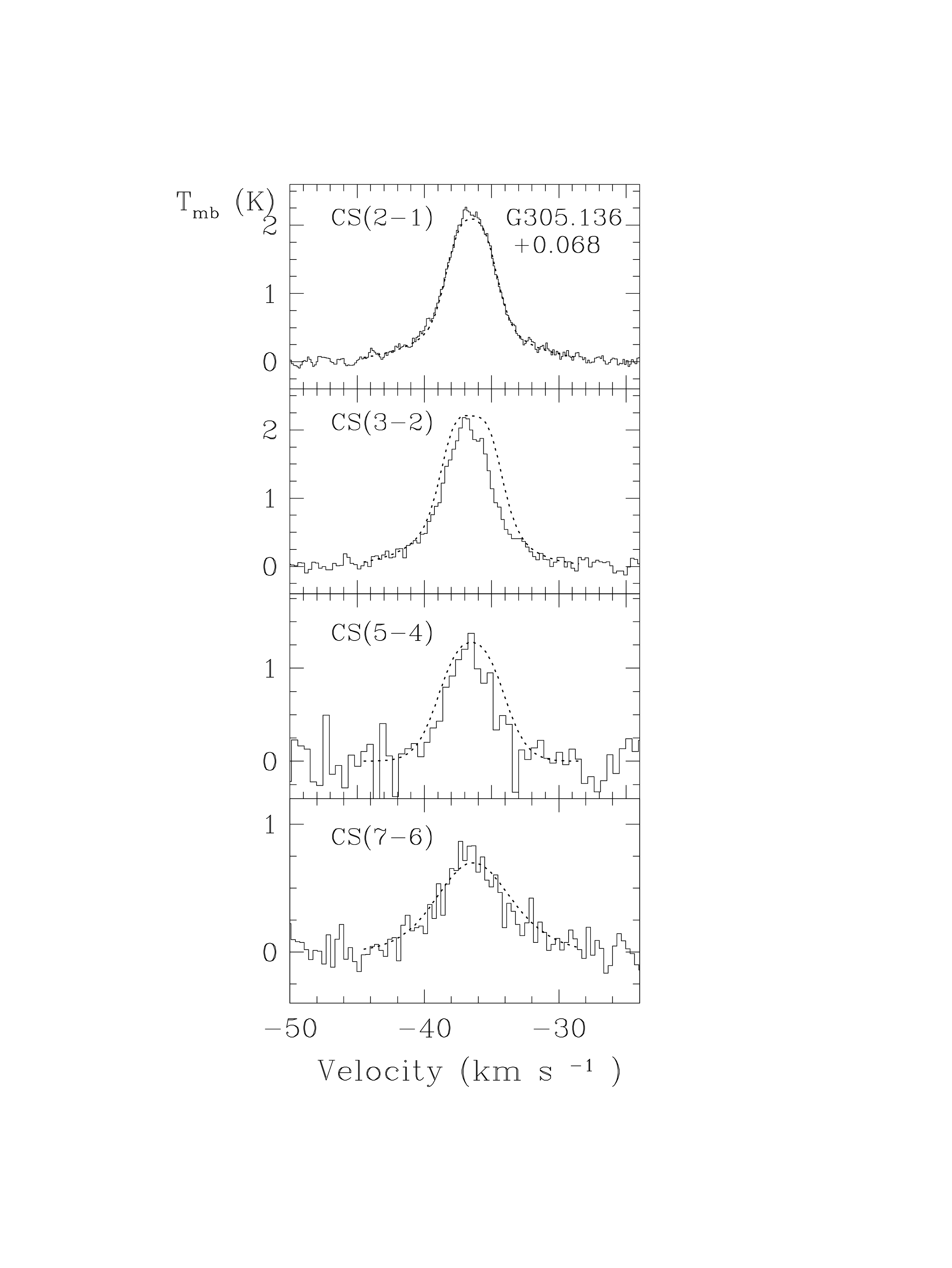}
\vspace{-30.0mm}
\caption
{Monte Carlo simulation of the CS transitions. Continuous lines correspond 
to the observed spectra and dotted lines to the simulated spectra.
\label{fig-montecarlo}}
\end{figure}

To model a cloud requires us to adopt the dependences of the physical 
parameters with radius. Here we assume that the density, kinetic temperature, 
velocity field and turbulent velocity follow power-law dependencies 
with radius.  Besides the great simplicity and flexibility in the program
provided by this choice, these are realistic assumptions. 
We find that the simple power-law model can very well fit the observed
data. This is illustrated in Figure~\ref{fig-montecarlo} 
which shows the four CS spectra observed at the peak position of the core 
and in dotted line the results of the best fit model.
We assumed a [CS/H$_2$] abundance ratio of $2.2\times10^{-10}$ \citep{Tafalla2004}.
The model fit implies that the core is very dense and highly centrally 
condensed, with the density depending with radius as, 
$$ n = 1.6\times10^5 \left(\frac{r}{0.3~{\rm pc}}\right)^{-1.8}~~ {\rm cm}^{-3}, $$ 
with minimum and maximum radius of 0.01 and 1 pc, respectively.
The model fit also indicates that the turbulent velocity increases toward 
the center of the core as,
$$ V_{turb} = 1.4 \left(\frac{r}{0.3~\rm{pc}}\right)^{-0.2}~~ \kms. $$
This result is supported by recent observations
which show that the linewith increases toward the center
(Servajean et al., private communication). Finally, the model fit 
also implies that the core is cold, and that the 
temperature slightly increases towards the center as,
$$ T_K= 16 \left(\frac{r}{0.3~{\rm pc}}\right)^{-0.15}~~ {\rm K}. $$


\subsection {Embedded sources}
\begin{deluxetable}{lccccccc}
\tablewidth{0pt}
\tablecolumns{8}
\tablecaption{EMBEDDED SPITZER SOURCES \label{tbl-spitzer}}
\tablehead{
\colhead{Source} & \colhead{RA(J2000)} & \colhead{Dec.(J2000)} & 
 \multicolumn{5}{c}{Flux density (mJy)} \\ 
\cline{4-8}
\colhead{} & \colhead{} & \colhead{}  & \colhead{3.6 $\mu$m}  & \colhead{4.5 $\mu$m} 
& \colhead{5.8 $\mu$m} & \colhead{8.0 $\mu$m} & \colhead{24.0 $\mu$m} 
}
\startdata
  A & 13 10 40.212 & $-62$ 43  9.03 & 0.8673 & 4.219 &  9.852 & 13.91 & 21.515 \\ 
  B & 13 10 42.713 & $-62$ 43 16.52 &  -$^a$ & 14.56 & 21.18  & 18.45 & 95.281 \\
  C & 13 10 44.158 & $-62$ 43 26.46 & 11.52  & 20.12 & 27.05  & 29.12 &  -$^b$ \\
\enddata
\tablenotetext{a}{Not given.}
\tablenotetext{b}{Projected towards a strong background extended emission.}
\end{deluxetable}

We searched for embedded sources within the core by inspecting all 
images available to the public taken with the {\it Spitzer} telescope.
Within a region of 20\arcsec\ in radius centered at the peak 
position of the dust core, we identified three compact sources that appear in 
both the 8 $\mu$m and the 24 $\mu$m images, making them good candidates 
for embedded objects. These sources, labeled A, B and C in 
Figure~\ref{fig-irac4}, were also detected in all the other IRAC bands.  
The observed parameters of the three objects are given 
in Table~\ref{tbl-spitzer}. Columns $2-3$ give the coordinates, columns $4-8$ 
the flux densities measured at 3.6, 4.6, 5.8, 8.0 and 24 $\mu$m, 
respectively.

The spectral energy distributions of these objects were 
analyzed by fitting model SEDs using a large grid of precomputed models 
\citep{Robitaille2007}. The SED fitting indicates that two of the three
sources (A and B) correspond to embedded objects. Source B, which is located 
closest to the peak position of the dust core, is most likely a 
very young protostellar object 
(age of $6\times10^4$ yr) with a mass of 2.6 \Msun, a luminosity of 66 \Lsun\ 
and associated with an envelope accreting at a high rate of $1.1\times10^{-4}$ 
\Msun\ yr$^{-1}$. Given its large envelope accretion rate, it is possible 
that object B is the seed of a high mass protostar. 
Observational support for the presence of infalling gas within the 
central region of the \gcoldcore\ cold core may be found in the 
observed profiles of the \cotd\ and \tcotd\ lines, the former one exhibiting
a broad asymmetric blueshifted profile and the later a narrower Gaussian 
profile.  Clearly, further observations are needed to investigate this hypothesis.
The SED fitting indicates that source A corresponds to a slightly 
older object (age of $2\times10^6$ yr), with a mass 
of 5.8 \Msun, a luminosity of $1.5\times10^3$ \Lsun, without an envelope but 
with an accretion disk. The SED fitting indicates that source C is not embedded, 
and that it has an age of $4\times10^6$ yr, a mass of 3.2 \Msun, a luminosity 
of 88 \Lsun\ and that is associated with a disk with an outer radius of 
80 AU.

\eject
\section{Summary}

We observed molecular line emission in the $J=3\rightarrow2$ transitions of CO 
and $^{13}$CO and in several transitions of CS toward the G305.136+0.068 cold 
dust core. Also observed was the 0.87 millimeter continuum emission.
The main results and conclusions are summarized as follows.

The molecular core is isolated and roughly circularly symmetric. 
However, the angular size (FWHM) of the core depends on the
transition observed, ranging from 18\arcsec\ in the \csss\ transition to
64\arcsec\ in the \cotd\ transition. This indicates that the core 
exhibits a steep density dependence with radius.
From the \cotd\ and \tcotd\ observations we derive an LTE mass of
$1.5\times10^3$ \Msun, in good agreement with the mass derived from 
the dust observations of $1.2\times10^3$ \Msun. 

Model fit of the CS emission using a Monte-Carlo radiative transfer 
code indicates that the 
\ccgtcc\ core is very dense and centrally condensed, with a density 
dependence with radius following a power law with an index of $-1.8$.
The model fit also indicates that the turbulent velocity and 
temperature increases toward the center of the core following 
power law with indices of $-0.2$ and $-0.15$, respectively.

A fit to the SED of the emission from millimeter to far infrared wavelengths
gives a dust temperature of 17 K, confirming that the core is cold, 
and an average column density of $2.6\times10^{23}$ cm$^{-2}$
or equivalently 1 gr cm$^{-2}$. This value is 
slightly above the theoretical threshold value required for the formation
of high mass stars \citep{Krumholz2008}.

We found two embedded sources within a region of $\sim 20$\arcsec\
centered at the core. The object closest to the core center is young, 
has a luminosity of 66 \Lsun\ and is accreting mass with a high accretion 
rate, of $\sim1\times10^{-4}$ \Msun\
yr$^{-1}$. This object may correspond to the seed of a high mass protostar
still in the process of formation. All the above results indicate that 
\ccgtcc\ corresponds to a massive and dense cold molecular core, in a very 
early evolutionary stage, in which the formation of a central massive object
may have just started.

\acknowledgements

G.G., Y.C. and D.M gratefully acknowledge support from CONICYT through 
project BASAL PFB-06. J.E.P. acknowledges funding from the European Community
Seventh Framework Programme (/FP7/2007-2013/) under grant agreement No 229517
and support by the Swiss National Science Foundation, project number
CRSII2-141880. A.E.G. acknowledges support from NASA Grants NNX12AI55G and 
NNX10AD68G. This publication made use of the GLIMPSE-Spitzer and 
Hi-GAL-Herschel database.

\newpage


\bibliographystyle{apj}
\bibliography{g305}



\end{document}